%% file: main.tex
\documentclass[sigconf]{acmart}

\usepackage{amssymb}          
\usepackage[noend]{algorithmic}
\usepackage{algorithm} 
\usepackage{multirow}         
\usepackage[table]{xcolor} 
\usepackage{colortbl}         
\usepackage{float}          
\usepackage{subfigure}        
\usepackage{xcolor}
\usepackage{enumitem}
\usepackage{bbold}

\setlist[itemize]{leftmargin=*}

\usepackage[T1]{fontenc}
\usepackage{soul}
\usepackage{color}

\definecolor{lightgray}{rgb}{0.75, 0.75, 0.75}
\sethlcolor{lightgray}

\usepackage[most]{tcolorbox}
\definecolor{codebg}{RGB}{245,245,245}
\newtcolorbox{promptbox}[2][]{
  enhanced,
  colback=codebg,
  colframe=black!75,
  coltitle=black,
  fonttitle=\bfseries\sffamily,
  title={#2},
  arc=2pt,
  boxrule=0.6pt,
  left=6pt,right=6pt,top=6pt,bottom=6pt,
  fontupper=\small\ttfamily\linespread{1.1}\selectfont, 
  #1
}

\usepackage{booktabs}
\usepackage{multirow}
\usepackage[table]{xcolor}

\definecolor{matchbg}{RGB}{236,245,236}      
\definecolor{mismatchbg}{RGB}{252,236,236}   
\definecolor{kw}{RGB}{90,90,90}

\AtBeginDocument{
  }

\def\new{\color{black}}

\copyrightyear{2026}
\acmYear{2026}
\setcopyright{cc}
\setcctype{by}
\acmConference[KDD 2026] {Proceedings of the 32nd ACM SIGKDD Conference on Knowledge Discovery and Data Mining V.2}{August 9--13, 2026}{Jeju Island, Republic of Korea.}
\acmBooktitle{Proceedings of the 32nd ACM SIGKDD Conference on Knowledge Discovery and Data Mining V.2 (KDD 2026), August 9--13, 2026, Jeju Island, Republic of Korea}
\acmISBN{979-8-4007-2259-2/2026/08}
\acmDOI{10.1145/3770855.3818169}

\begin{document}

\title{LLM-as-a-Judge for Reliable and Explainable Offline Evaluation in Top-K Recommendation}

\author{Yue Que}
\email{yueque2-c@my.cityu.edu.hk}
\orcid{0009-0001-0468-6631}
\affiliation{
  \institution{City University of Hong Kong}
  \city{Hong Kong}
  \country{China}
}
\author{Junyi Zhou}
\email{junyizhou8-c@my.cityu.edu.hk}
\orcid{0009-0008-9616-4240}
\affiliation{
  \institution{City University of Hong Kong}
  \city{Hong Kong}
  \country{China}
}
\author{Xiaokun Zhang}
\authornote{Corresponding authors}
\email{dawnkun1993@gmail.com}
\orcid{0000-0002-9755-2471}
\affiliation{
  \institution{City University of Hong Kong}
  \city{Hong Kong}
  \country{China}
}
\author{Haiming Jin}
\email{jinhaiming@sjtu.edu.cn}
\orcid{0000-0001-5178-7198}
\affiliation{
  \institution{Shanghai Jiao Tong University}
  \city{Shanghai}
  \country{China}
}
\author{Qiao Xiang}
\email{xiangq27@gmail.com}
\orcid{0000-0002-3394-6279}
\affiliation{
  \institution{Xiamen University}
  \city{Xiamen}
  \country{China}
}
\author{Chen Ma}
\authornotemark[1]
\email{chenma@cityu.edu.hk}
\orcid{0000-0001-7933-9813}
\affiliation{
  \institution{City University of Hong Kong}
  \city{Hong Kong}
  \country{China}
}

\renewcommand{\shortauthors}{Yue Que et al.}

\input{abstract}




\maketitle

\newcommand\kddavailabilityurl{https://doi.org/10.5281/zenodo.20283783}
\ifdefempty{\kddavailabilityurl}{}{
\begingroup\small\noindent\raggedright\textbf{Resource Availability:}\\
The source code of this paper has been made publicly available at \url{\kddavailabilityurl}.
\endgroup
}

\input{introduction}
\input{method}

\input{experiment}
\input{related}
\input{conclusion}

\begin{acks}
This work is supported by the Early Career Scheme (No.CityU 21219323) and the General Research Fund (No.CityU 11220324) of the University Grants Committee (UGC), the NSFC Young Scientists Fund (No.9240127), and the Donation for Research Projects (No.9229164 and No.9229216). 
\end{acks}

\balance

\bibliographystyle{ACM-Reference-Format}
\bibliography{ref}
\appendix
\input{appendix}
\end{document}

%% file: abstract.tex
\begin{abstract}
Recommendation evaluation plays a crucial role in guiding the refinement and deployment of  recommender systems. 
Most existing trials rely on offline evaluation using Top-K metrics computed over holdout user behaviors. 
However, we identify two fundamental limitations that undermine their ability to deliver reliable and explainable evaluations. 
Regarding reliability, offline evaluation treats observed user feedback as a proxy of true preferences and enforces rigid ID matching between the proxy and recommendation.
In practice, feedback collections are inherently shaped by incomplete and biased item exposure, leading to distorted and unreliable assessments.
Regarding explainability, Top-K metrics only establish numerical scores without offering meaningful insights to support them, thereby reinforcing the black-box nature of offline evaluation.

In this paper, we propose a reliable and explainable LLM-as-a-Judge framework for offline  recommendation evaluation.
To enhance reliability, we introduce a semantic proxy from user textual behaviors to represent their true preferences.
This proxy allows for more flexible matching between preferences and recommendations in the semantic space, rather than depending on the holdout feedback.
To ensure explainability, the LLM Judge adopts a \textit{reasoning-then-scoring} process to generate relevance judgments along with explicit rationale.
Finally, we aggregate the individual scores into global Top-K metrics to quantify overall recommendation quality, and provide justification for each preference hit or miss.
Extensive experiments demonstrate that the LLM Judge achieves solid reliability, explainability, and robustness in evaluation.
\end{abstract}

%% file: introduction.tex
\section{Introduction}
Recommender systems enhance user experience by providing personalized recommendation to alleviate information overload. 
Evaluating recommendation performance is essential for the system design, refinement and deployment.
Serving as the widely adopted paradigm in research, offline evaluation assesses recommendation quality by computing common Top-K metrics such as hitting rate, Recall@K, or NDCG@K on holdout feedback collections \cite{holdout, eval_rs}.

Unfortunately, under the widespread presence of missing-not-at-random (MNAR) feedback, offline metrics risk being distorted and thus provide unreliable evaluation results.
Ideally, the gold standard for recommendation assessment can be defined as: \textit{to match true preference space of the user}.
Offline evaluation simply assumes that collected user feedback is the proxy for the entire preference space.
Based on this assumption, offline metrics reflect recommendation quality by performing strict ID matching between the recommended items and the test set.
While in practice, user history collections are largely MNAR, which means they only represent a subset of the overall interests due to multiple sources of inherent bias, such as exposure control imposed by the deployed system or user item-viewing behaviors \cite{closed_loop}.
Consequently, enforcing rigid ID matching on MNAR data fails to account for unobserved interests \cite{kun1}, resulting in an inevitable deviation from the gold standard.

On the other hand, offline evaluation remains limited in delivering explainable assessments \cite{podcast}.
With growing attention on explainable recommendation, explainability in evaluation is also emphasized to offer meaningful insights into assessment rationality \cite{explan, why}.
However, traditional offline metrics are typically confined to establishing numerical performance benchmarks and lack the ability to provide sufficient explanations.
Based on these observations, we highlight the fundamental limitation in current offline evaluation: \textit{the absence of reliable and explainable proxy that consistently aligns with the gold standard}.

To move beyond the feedback collection proxy employed in offline evaluation, Large Language Models (LLMs) demonstrate the capability to infer a semantic proxy of the gold standard.
With extensive world knowledge and excellent reasoning ability, LLMs can interpret user behavioral histories to capture latent intentions and unobserved interests beyond MNAR feedback \cite{kun2}.
Therefore, the semantic representation offers a more comprehensive summary of user true preferences, enhancing the reliability of the preference proxy. 
Building on this, the emergence of LLM-as-a-Judge paradigm further promotes explainable evaluation by engaging agents to reason over candidates and align judgments with human-like criteria \cite{laaj_survey, llm_rele_judge}.
Taken together, our approach leverages the semantic understanding and explainability of LLMs to effectively bridge the gap toward reliable and explainable evaluation.

In this paper, we propose a novel LLM-as-a-Judge framework to address the fundamental limitations in offline recommendation evaluation.
First, to enhance the reliability in offline evaluation, we introduce a semantic proxy to represent user preference space, thus moving beyond the reliance on supplementary test sets.
Based on this proxy, we establish a semantic matching principle between the proxy and the recommended content, enabling more flexible and logical alignment with user interests.
Second, to improve the explainability in offline evaluation, we devise a two-step judging process within the LLM Judge.
In the reasoning step, the Judge summarizes user intentions from the history to build the semantic proxy, and assesses the relevance between the proxy and recommendation.
In the scoring step, the Judge assigns pointwise scores to candidates based on the prior reasoning and outputs explicit rationales.
Finally, the judgments are aggregated into global metrics that reflect overall recommendation performance, accompanied by individual reasons explaining each decision.

In summary, our contributions are threefold.
\begin{itemize}
    \item We propose an LLM-as-a-Judge framework that introduces a semantic proxy to represent user true preferences. This framework enables flexible semantic matching between the proxy and recommended content, providing reliable evaluation.
    \item We devise a reasoning-then-scoring judging process in the LLM Judge to produce assessment scores along with explicit rationales, achieving explainable evaluation.
    \item {\new We conduct extensive experiments to demonstrate the reliability and explainability of the Judge, as well as the evaluation robustness across varying LLM model scales and input configurations.}
\end{itemize}

%% file: method.tex
\section{Methodology}
In this section, we first outline the preliminaries of Top-K recommendation and common offline metrics to formulate the offline evaluation problem.
Next, we introduce the proposed LLM-as-a-Judge framework for recommendation evaluation.

\begin{figure*}[t]
  \centering
  \includegraphics[width=1.0\textwidth]{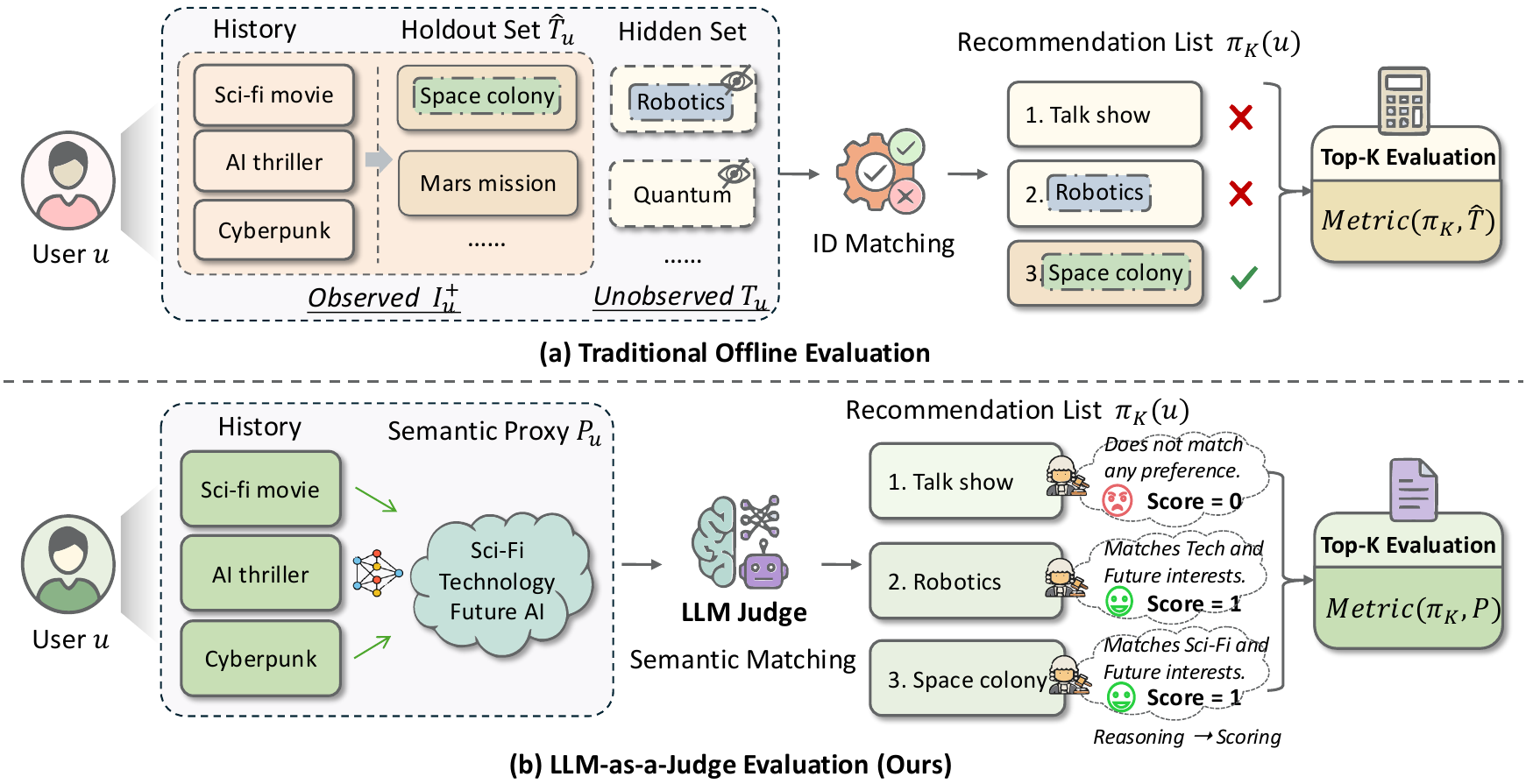}
  \caption{
    \textbf{Overview of traditional offline evaluation vs. LLM-as-a-Judge evaluation in Top-K recommendation.}
  }
  \label{fig:method_overview}
  \Description{method}
\end{figure*}

\subsection{Problem Formulation}
We study the standard Top-K recommendation task in collaborative filtering, which serves as a fundamental pipeline for building recommender systems \cite{cf}.
Let $\mathcal{U}$ and $\mathcal{I}$ denote the user and item sets.
The collection of user-item interactions is represented as $\mathcal{D}=\{(u,i)\mid u\in\mathcal{U}, i\in\mathcal{I}\}$, where each pair  $(u,i)\in\mathcal{D}$ indicates that user $u$ has interacted with item $i$.
For each user $u$, we denote his historical feedback set as $\mathcal{I}^+_u=\{i\in\mathcal{I}\mid (u,i)\in\mathcal{D}\}$.

Ideally, a collaborative filtering recommender is trained on the observed interactions $\mathcal{I}^+_u$ to learn a scoring function $s(u,i)$ that estimates a user’s preference for each item.
During inference, the model computes scores for user $u$ over the set of unseen items $\mathcal{I}\setminus\mathcal{I}^+_u$, and ranks them in descending order of $s(u,i)$ to produce the Top-K recommendation list:
\begin{equation}
\pi_K(u)=\operatorname{Top-K}_{i\in\mathcal{I}\setminus\mathcal{I}^+_u} \, s(u,i),
\end{equation}
where $\pi_K(u)=\{i_1,\ldots,i_K\}$ denotes the ordered list of the $K$ highest-scoring items.
The recommendation model is evaluated by comparing $\pi_K(u)$ with the ground-truth relevant items from the unseen set.
We define the set of items that truly match user $u$'s preference as $\mathcal{T}_u\subseteq\mathcal{I}\setminus\mathcal{I}^+_u$.

To evaluate how effectively $\pi_K$ captures ground-truth preference in $\mathcal{T}$, two widely used offline metrics are employed, which are the recall rate and the normalized discounted cumulative gain (NDCG). 
Recall@$K$ measures the fraction of relevant items that appear in the Top-K recommendation list:
\begin{equation}
\mathrm{Recall}@K=\frac{1}{|\mathcal{U}|}\sum_{u\in\mathcal{U}}{\frac{|\pi_K(u)\cap \mathcal{T}_u|}{|\mathcal{T}_u|}}.
\end{equation}
NDCG@$K$ evaluates both the relevance and ranking quality of recommendations by assigning more scores to correct recommendations with higher rankings.
The discounted cumulative gain (DCG) is first calculated as:
\begin{equation}
\mathrm{DCG}@K(u)=\sum_{r=1}^{K}\frac{2^{\mathbb{1}\left\{i_r\in\mathcal{T}_u|i_r\in\pi_K(u)\right\}}-1}{\log_2(r+1)}.
\end{equation}
Then ideal DCG is computed by re-ranking hitted items to the top:
\begin{equation}
\mathrm{iDCG}@K(u)=\sum_{r=1}^{\min(K,|\mathcal{T}_u|)}\frac{1}{\log_2(r+1)}.
\end{equation}
Finally, NDCG is obtained based on the ratio of DCG to iDCG:
\begin{equation}
\mathrm{NDCG}@K=\frac{1}{|\mathcal{U}|}\sum_{u\in\mathcal{U}}\frac{\mathrm{DCG}@K(u)}{\mathrm{iDCG}@K(u)}.
\end{equation}

In practice, as illustrated in Figure~\ref{fig:method_overview} (a), offline evaluation typically uses a holdout subset $\hat{\mathcal{T}}$ from the observed $\mathcal{I}^+$ as a proxy for the unobserved true preference $\mathcal{T}$.
Traditional Top-K metrics measure performance through strict ID matching between $\pi_K$ and $\hat{\mathcal{T}}$, \textit{i.e.,} $\mathrm{Metric}(\pi_K,\hat{\mathcal{T}})$, thereby resulting in deviation from the ideal gold standard.
To address this issue, we formulate the offline evaluation problem as identifying a proxy $\mathcal{P}$ that yields evaluation results (\textit{e.g.,} Recall@$K$ and NDCG@$K$) consistent with those computed on $\mathcal{T}$, that is, $\mathrm{Metric}(\pi_K,\mathcal{P})\rightarrow\mathrm{Metric}(\pi_K,\mathcal{T})$.

\subsection{LLM-as-a-Judge}
We introduce the LLM Judge framework in three parts:
(1)\textit{ Judging Process}, which outlines the procedures for performing pointwise semantic matching and reasoning; (2) \textit{Prompt Construction}, which defines concrete principles to effectively guide the agent to judge; and (3) \textit{Outcome Evaluation}, which aggregates the generated judgments into standard Top‑K performance metrics.

\subsubsection{Judging Process}
We assume that each item is associated with side information \cite{kun3} in natural language (\textit{e.g.,} title, attributes, or short description).
As shown in Figure~\ref{fig:method_overview} (b), given the user historical interactions and a Top-K recommendation list produced by the recommender, the Judge is prompted to assess the pointwise relevance between the user and each candidate item based on their linguistic features.
Central to our LLM Judge is the semantic matching mechanism, which replaces the traditional ID-based matching employed in holdout test evaluations.
For clarity, we conceptually decompose the judgment into \emph{reasoning} and \emph{scoring} steps, while practically encapsulating both within a single LLM query.


In the \emph{reasoning step}, the Judge first summarizes user preferences from the observed textual feedback, forming the semantic proxy that captures their high-level interests.
Based on this proxy, the Judge performs semantic matching to identify the relevance between recommendation candidates and the inferred user profile.
This semantic-level inference allows the model to rely on linguistic information rather than meaningless item IDs, thereby effectively alleviating the MNAR impact inherent in traditional holdout proxies.
For each candidate item, the matching judgment involves chain-of-thought reasoning to cite key evidence that leads to an accept or reject decision. 
The reasoning outcome is a single-sentence rationale explaining why the item aligns (or mismatches) the underlying preferences.
This stage guides the Judge to ground its assessment on observable textual cues, instead of directly guessing the score.


In the \emph{scoring step}, the Judge assigns a pointwise relevance score to each candidate item conditioned on the preceding reasoning rationale.
The score indicates whether the candidate is relevant to the inferred user preferences (1 for relevant, 0 for irrelevant).
Along with the score, rationale of the judgment is also reported to ensure transparent and interpretable evaluation outcome. 


\begin{figure}[t]
\centering
\begin{minipage}{\columnwidth}
\begin{promptbox}{\textcolor{white}{Prompt Template}}

\textbf{\underline{{\fontfamily{fvm}\itshape\footnotesize SYSTEM PROMPT}}}

You are a strict and impartial recommendation judge for \hl{[INPUT: TASK DEFINITION]}.

Your goal is to determine if a candidate item is a perfect match for the user based on user history.

\textbf{BASIC PRINCIPLES}

1. Reasoning-then-scoring: First infer user preferences from the history, then think about the reason that leads to matching or mismatching, finally make your judgment based on the reasoning. 

2. Reject-by-default: Your starting position is REJECT (0). Remember most recommendations are mediocre, you only approve it if the evidence is undeniable.

\textbf{DECISION FRAMEWORK}

0 : REJECT (Default)

\qquad - \hl{[INPUT: MATCHING CRITERIA]}

1 : HIT (Strict Match)

\qquad - \hl{[INPUT: MATCHING CRITERIA]}

\textbf{OUTPUT FORMAT}

You MUST output a JSON object with 'reason' and 'score' fields. The 'reason' should be a single, concise sentence in English.

\vspace{\baselineskip}
\textbf{\underline{{\fontfamily{fvm}\itshape\footnotesize USER PROMPT}}}

User history: \hl{[INPUT: HISTORY]}

Candidate item to evaluation: \hl{[INPUT: CANDIDATE]}

\end{promptbox}
\end{minipage}
\caption{Evaluation Prompt applied in our LLM Judge. Basic principles and output format highlighted in boldface are fixed regions, while the input fields shaded in gray can be customized for different tasks.}
\label{fig:judge-prompt}
\Description{prompt}
\end{figure}

\subsubsection{Prompt Construction}
Our prompt is designed to implement a strict pointwise relevance evaluator.
To ensure generalizability, the prompt template standardizes only the general evaluation procedure, while task-specific details are provided through customizable input fields to maintain scalability across various recommendation tasks. 
As shown in Figure \ref{fig:judge-prompt}, the prompt template explicitly specifies three key aspects in the evaluation process: \emph{basic principles}, \emph{input fields}, and \emph{output format}.

\emph{Basic principles.}
We define two fundamental rules to guide the entire judgment process, (1) the reasoning-then-scoring principle, and (2) the reject-by-default policy.
The first rule regulates the Judge to follow our proposed two-step judgment procedure, while the reject-by-default policy enables a conservative LLM Judge to mitigate the over-approve tendency in LLMs.
Given that LLMs often pursue divergent associations and may exhibit hallucinations \cite{Hallucination}, we explicitly constrain the Judge to always reject the candidate unless the provided evidence clearly supports a perfect match.
This rule is therefore essential for filtering out ambiguous candidates and ensuring an accurate evaluation process.

\emph{Input fields.}
The input fields are serialized into a unified template consisting of (1) an instruction block that defines the task and matching criteria, (2) the textual fields representing the user historical interactions and the recommended candidate information.
In the instruction block, the current recommendation task requires customized definition.
The matching criteria are originally task-agnostic with a general description of the perfect matching situation, but we also allow flexible customization to better fit the corresponding task.
This modular design allows the same judging logic to transfer across different recommendation scenarios by simply adjusting the instruction block while keeping the remaining template and rubric unchanged.

\emph{Output format.}
Lastly, the Judge is required to output a binary relevance score (0 or 1) along with a single-sentence rationale explaining the judgment.
The binary scores are then aggregated into global metrics to reflect the overall recommendation quality, while their corresponding explanations are retained for further qualitative analysis or error diagnosis to better understand the Judge's behaviors and reasoning patterns.

\subsubsection{Outcome Evaluation}

Since the judgment score takes binary value, we follow typical offline evaluation protocol to aggregate the LLM evaluation outcomes with standard Top-K recommendation metrics.
In this study, we select the representative Recall@$K$ and NDCG@$K$ as primary evaluation metrics, while other list‑level metrics remain compatible with our framework.
We make minor adjustments to adapt the two metrics to our evaluation framework. 


Formally, given the Top-K list $\pi_K(u)=\{i_1,\ldots,i_K\}$, and the LLM judge outcome $y_{u,i_r}\in\{0,1\}$ for each rank $r$.
$\mathrm{NDCG}_{\text{Judge}}@K$ is computed in the standard way by treating $y_{u,i_r}$ as the relevance signal in DCG.
For Recall@$K$, our protocol does not rely on the holdout test set to represent $\mathcal{T}_u$, and thus $|\mathcal{T}_u|$ is unavailable.
Accordingly, we normalize the number of judged hits by $K$:
\begin{equation}
\mathrm{Recall}_{\text{Judge}}@K=\frac{1}{|\mathcal{U}|}\sum_{u\in\mathcal{U}}\frac{1}{K}\sum_{r=1}^{K} y_{u,i_r}.
\end{equation}

%% file: experiment.tex
\section{Experiment}

\begin{table*}[t!]
    \centering
    \caption{Consistency performance of normal testing and LLM Judge evaluation. The best consistency results are highlighted in boldface. Gain denotes the performance improvement over normal testing. Superscript {\Large *} indicates the statistically significant consistency with the ground truth (\textit{i.e.}, two-sided t-test with $\mathbf{p<0.05}$).
    }
    \label{tab:main}
    \resizebox{\textwidth}{!}{
        \begin{tabular}{c c c c c c c c c c c } 
            \toprule
             Dataset&Metric& Normal Testing&LLM Judge&Gain& Normal Testing&LLM Judge&Gain& Normal Testing&LLM Judge&Gain\\
             \midrule
             \multirow{8}*{Kuairec}
             &Top-K&\multicolumn{3}{c}{Recall@20}& \multicolumn{3}{c}{Recall@50}& \multicolumn{3}{c}{Recall@100}\\
             \cmidrule(lr){2-11}
             &Pearson Correlation& 35.56\%& \textbf{87.62\%*}& +52.07\%& 31.93\%& \textbf{79.79\%*}& +47.86\%& 27.88\%& \textbf{76.71\%*}& +48.83\%\\
             &Spearman's Rank& 26.19\%& \textbf{85.71\%*}& +59.52\%& 19.05\%& \textbf{64.29\%}& +45.24\%& 16.67\%& \textbf{79.42\%*}& +62.75\%\\
             &Kendall's $\tau$& 28.57\%& \textbf{71.43\%*}& +42.86\%& 28.57\%& \textbf{42.86\%}& +14.29\%& 21.43\%& \textbf{61.83\%*}& +40.40\%\\
            \cmidrule(lr){2-11}
             &Top-K&\multicolumn{3}{c}{NDCG@20}& \multicolumn{3}{c}{NDCG@50}& \multicolumn{3}{c}{NDCG@100}\\
             \cmidrule(lr){2-11}
             &Pearson Correlation& 2.65\%& \textbf{62.61\%}& +59.96\%& -2.88\%& \textbf{70.79\%*}& +73.66\%& 2.43\%& \textbf{71.35\%*}& +68.92\%\\
             &Spearman's Rank& 11.90\%& \textbf{71.43\%*}& +59.52\%& 16.67\%& \textbf{78.57\%*}& +61.90\%& 16.67\%& \textbf{80.95\%*}& +64.29\%\\
             &Kendall's $\tau$& 14.29\%& \textbf{50.00\%}& +35.71\%& 21.43\%& \textbf{64.29\%*}& +42.86\%& 21.43\%& \textbf{64.29\%*}& +42.86\%\\
            \midrule
             \multirow{8}*{Coat}
             &Top-K&\multicolumn{3}{c}{Recall@3}& \multicolumn{3}{c}{Recall@5}& \multicolumn{3}{c}{Recall@10}\\
             \cmidrule(lr){2-11}
             &Pearson Correlation& 38.15\%& \textbf{87.41\%*}& +49.25\%& 34.06\%& \textbf{85.70\%*}& +51.64\%& -8.91\%& \textbf{76.20\%*}& +85.11\%\\
             &Spearman's Rank& 23.81\%& \textbf{92.86\%*}& +69.05\%& 9.52\%& \textbf{52.41\%}& +42.89\%& -16.67\%& \textbf{38.10\%}& +54.76\%\\
             &Kendall's $\tau$& 21.43\%& \textbf{78.57\%*}& +57.14\%& 7.14\%& \textbf{40.01\%}& +32.86\%& -7.14\%& \textbf{21.43\%}& +28.57\%\\
            \cmidrule(lr){2-11}
             &Top-K&\multicolumn{3}{c}{NDCG@3}& \multicolumn{3}{c}{NDCG@5}& \multicolumn{3}{c}{NDCG@10}\\
             \cmidrule(lr){2-11}
             &Pearson Correlation& -77.57\%& \textbf{73.86\%*}& +151.43\%& -48.31\%& \textbf{85.61\%*}& +133.93\%& -30.12\%& \textbf{82.72\%*}& +112.85\%\\
             &Spearman's Rank& -76.19\%& \textbf{80.95\%*}& +157.14\%& -47.62\%& \textbf{76.19\%*}& +123.81\%& -45.24\%& \textbf{83.33\%*}& +128.57\%\\
             &Kendall's $\tau$& -57.14\%& \textbf{64.29\%*}& +121.43\%& -28.57\%& \textbf{57.14\%*}& +85.71\%& -35.71\%& \textbf{71.43\%*}& +107.14\%\\
        \bottomrule
        \end{tabular}
    }
\end{table*}

In this section, we conduct extensive experiments to validate the effectiveness of our LLM Judge in achieving reliable and explainable evaluation for Top-K recommendation.
We aim to answer the following research questions:
\begin{itemize}
    \item \textbf{RQ1:} Can the LLM Judge be a reliable proxy for the unbiased offline evaluation?
    \item \textbf{RQ2:} Can the LLM Judge generate logically sound and persuasive explanations for its judgment?
    \item \textbf{RQ3:} {\new How robust are the evaluation results under different LLM model scales and input configurations?}
\end{itemize}

\subsection{Evaluation Pipeline}
\subsubsection{Data Overview}
We employ two real-world datasets, Kuai-
rec\footnote{\url{https://kuairec.com/}} \cite{kuairec} and Coat\footnote{\url{https://www.cs.cornell.edu/~schnabts/mnar/}} \cite{ips2}, both of which are widely used for unbiased evaluation in recommender systems. 
Kuairec is a video recommendation dataset containing 12.5 million interactions between 7.1k users and 10.7k items.
Coat dataset records 6.5k ratings provided by 290 users on 300 clothes from a shopping platform.
For model training, both datasets adopt the same MNAR feedback as most recommendation datasets.
While for testing, they support unbiased evaluation with dedicated test sets.
Specifically, Kuairec provides a fully observed test subset containing 1.4k users and 3.3k items, where each user has left feedback on every item in the subset.
Coat offers an unbiased test set in which all users have rated 16 randomly exposed items.
Among available real-world datasets, only Kuairec, Coat, and Yahoo!R3 \cite{yahoo} provide unbiased offline test data. 
We exclude Yahoo!R3 as it contains only item IDs without textual information, thus being inapplicable to our setting.
We provide the textual information details of Kuairec and Coat in Appendix~\ref{app_input}.

\subsubsection{Baseline Construction}
We design two pipelines to establish the lower and upper bounds of evaluation quality, namely normal testing and unbiased testing.
We split the MNAR feedback of each dataset into training and validation sets with a 7:3 ratio, and train recommendation models on the training set.
In normal testing, model performances are evaluated by the validation set, which represents a lower bound of evaluation quality due to incomplete item exposure.
In contrast, unbiased testing assesses the same trained models on the unbiased test sets, which are collected under full item exposure, thereby serving as the evaluation upper bound.
To enable unbiased testing, we binarize the fully exposed test data into positive and negative samples, defining positives as videos watched at least twice in Kuairec and ratings above 3 in Coat dataset.
We then treat the unbiased testing as the ground truth, and consider both normal testing and LLM Judge as proxy evaluation schemes that approximate this ground truth.
Among them, normal testing serves as the baseline for comparison with the LLM Judge.
Finally, to characterize the evaluation landscapes of each proxy and the ground truth, we instantiate them as performance ranking lists of recommendation models derived from their respective evaluations.

\subsection{Experimental Settings}
\subsubsection{Recommendation Models}
To build the model ranking list, we implement a diverse set of representative collaborative filtering models.
They include matrix factorization methods: MF \cite{mf} and NeuMF \cite{neumf}; graph neural network models: NGCF \cite{ngcf}, LR-GCCF \cite{lrgccf}, and LightGCN \cite{lightgcn}; contrastive learning model: SimGCL \cite{simgcl}, and graph-based variants that improve high-order connectivity modeling: MGCCF \cite{mgccf}, GTN \cite{gtn}, and CAGED \cite{caged}.
Rather than identifying an optimal recommender in the list, we focus on the relative model rankings under the two proxy schemes, and assess their consistencies with the unbiased ordering to reflect evaluation reliability.
We attach model implementation details in Appendix~\ref{recmodel}.

\begin{figure*}[t!]
  \centering
  \subfigure[Kuairec]{\includegraphics[width=\linewidth]{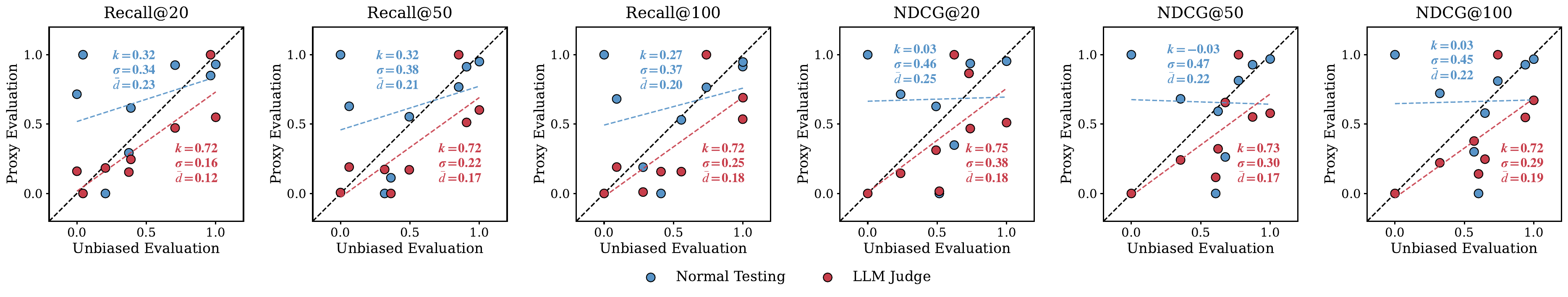}}
  \subfigure[Coat]{\includegraphics[width=\linewidth]{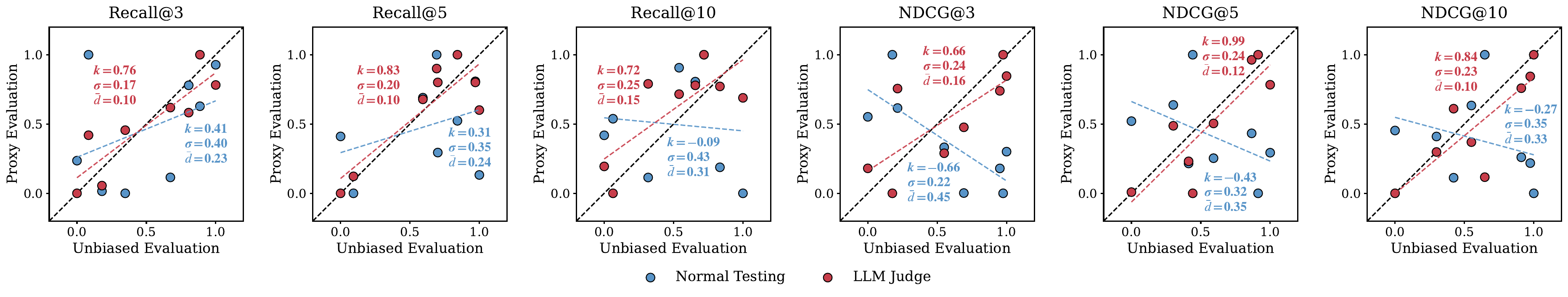}}
  \caption{Scatter plot between proxy evaluation and unbiased evaluation. Diagonal line $\boldsymbol{y=x}$ represents absolute absence of discrepancy between two evaluation types. Linear regression statistics (${\boldsymbol{k}}$ denotes slope, $\boldsymbol{\sigma}$ represents standard deviation, and $\boldsymbol{\bar{d}}$ indicates MAE) of normal testing and LLM Judge are plotted to illustrate the accuracy trend.}
  \Description{linear}
  \label{fig:linear}
\end{figure*}

\subsubsection{Evaluation Metrics} \label{metric}
We employ a variety of evaluation metrics to examine the reliability and explainability of our LLM Judge separately.
To evaluate the reliability of proxy evaluation, we leverage four statistical metrics as follows:
\begin{itemize}
    \item \textbf{Pearson Correlation:} measures the consistency of score gaps between neighboring ranks.
    \item \textbf{Spearman's Rank:} evaluates the consistency of overall ordering positions between two ranked lists, irrespective of the score scale.
    \item \textbf{Kendall's $\boldsymbol{\tau}$:} quantifies the consistency of pairwise order relations between all entry pairs across the two lists.
    \item \textbf{Mean Absolute Error (MAE):} computes the average absolute difference between the proxy and unbiased evaluation scores after normalization.
\end{itemize}
Among these, the first three are correlation-based metrics taking values within $[-1, 1]$, where $1$ indicates perfect consistency, $0$ denotes no correlation, and $-1$ reflects complete inverse correlation. 
In contrast, MAE serves as an accuracy-based metric that captures the magnitude of deviation between two normalized evaluation scores.
Since we summarize recommendation performances in Top-K metrics (\textit{i.e.,} Recall@$K$ and NDCG@$K$), for comprehensive comparison, evaluations are conducted on recommendation lists with various lengths $K$. 
According to the sample size in each dataset, we employ $K\in\{20,50,100\}$ in Kuairec, and $K\in\{3,5,10\}$ in Coat.

To assess the explainability of the LLM Judge, we design explanatory rubrics from three complementary dimensions:
\begin{itemize}
    \item \textbf{Coherence:} whether the reasoning presented by the LLM Judge follows a logically consistent chain of thought that justifies its final decision.
    \item \textbf{Faithfulness:} whether the evidence mentioned in the explanation is genuinely supported by historical data rather than being hallucinated.
    \item \textbf{Persuasiveness:} whether the rationale can convince human evaluators to its judgment.
\end{itemize}

\subsubsection{LLM Judge Backbones}
To examine the evaluation robustness of the LLM Judge, we experiment with a series of LLMs with varying scales to build the Judge, including Qwen3-4B-Instruct-2507 \cite{qwen}, Qwen3-8B, Qwen3-14B, Qwen3-30B-A3B, and DeepSeek-V3.2 \cite{ds}.
As we mainly focus on investigating the effect of LLM capacity on the Judge performance, we do not introduce more models from other LLM families for comparison.
In the main experiments, Qwen3-30B-A3B is adopted as the default backbone to balance evaluation performance and computational cost. 
We report its runtime overhead in Appendix~\ref{overhead}.
For all backbones, the temperature is fixed at 0 to ensure deterministic outputs.

\subsection{Study of Evaluation Reliability (RQ1)}
To validate the reliability of the LLM Judge as a proxy for unbiased offline evaluation, we compare the Judge performance with normal testing from two perspectives.
From a relative ranking perspective, we investigate whether the Judge can accurately recover the relative performance order of recommendation models by measuring the ranking consistency between proxy evaluation and unbiased testing across multiple Top-K metrics.
From an absolute score perspective, we examine whether the Judge can produce consistent performance scores by directly comparing the normalized proxy scores with the unbiased scores at the model level and analyzing their overall trends.
We present detailed observations below. 

\emph{Ranking analysis.}
In Table~\ref{tab:main}, we report the ranking consistency between each proxy evaluation scheme and the unbiased testing under different Top-K metrics, measured by Pearson correlation, Spearman's rank, and Kendall's $\tau$.
On both datasets, normal testing exhibits low consistencies across all Top-K metrics, indicating that MNAR feedback substantially distorts both the relative rankings and the score gaps among recommendation models.
As a result, normal testing fails to align with unbiased testing, leading to unreliable performance rankings. 
In contrast, the LLM Judge consistently achieves strong correlations across all Top-K metrics, showing solid superiority over normal testing in maintaining ranking consistency with the ground truth.
Notably, on the NDCG evaluations of Coat, the Judge not only improves the consistency values but also reverses strong negative correlations of normal testing into positive agreement.
Furthermore, significance tests confirm that the consistency achieved by the Judge is statistically significant rather than incidental.
Therefore, the LLM Judge demonstrates promising capability to recover the relative model rankings much closer to the unbiased testing results.

\emph{Score analysis.}
Figure~\ref{fig:linear} illustrates the score-level agreement between proxy evaluations and unbiased testing, where each point represents a recommendation model with normalized performance scores in $[0,1]$.
We perform regression analysis for both normal testing and LLM Judge, where the slope $k$ reflects how well each proxy preserves relative score gaps (ideal value $k=1$) and the standard deviation $\sigma$ indicates evaluation stability.
$\bar{d}$, representing the MAE value, is computed as the average distance of points from the diagonal $y=x$ to quantify the absolute discrepancy from unbiased scores.
In normal testing, points are widely dispersed with noticeable deviation from the diagonal line, exhibiting substantial bias in score alignment.
This pattern is further supported by the regression statistics, where the slope $k$ generally deviates from the ideal value $1$, and the standard deviation $\sigma$ indicates greater instability in evaluation.
Additionally, the MAE result $\bar{d}$ reveals larger differences from the ground-truth scores.
In contrast, under the LLM Judge evaluation, performance points cluster more closely around the diagonal, demonstrating a clearer linear trend. 
Correspondingly, the slopes approach $1$ with both lower $\sigma$ and $\bar{d}$, suggesting that the Judge not only better preserves the relative score differences among models but also provides more accurate absolute performance estimates.

These findings confirm that the LLM Judge attains both ranking consistency and score-level agreement with unbiased testing.
Consequently, the Judge effectively improves the reliability of offline evaluation over normal testing.

\subsection{Study of Evaluation Explainability (RQ2)}
To investigate the evaluation explainability of the LLM Judge, we analyze the judgment rationales with both quantitative and qualitative approaches.
In the quantitative analysis, we invite human experts to assess the quality of the rationales along the three key dimensions: coherence, faithfulness, and persuasiveness (\textit{cf.} section~\ref{metric}).
In the qualitative analysis, we further conduct the case study on rationale patterns to explore the types of reasoning that the Judge tends to produce.

\emph{Rationale quality assessment.}
We first categorize the Judge outcomes into \textit{Match} and \textit{Mismatch} groups based on the Judge decision, as the two types of reasons differ significantly.
From each group, we randomly sample 100 judgment rationales, and engage three human evaluators to manually annotate each rationale with binary scores across the three dimensions.
The average human evaluation results are summarized in Table~\ref{tab:human}.
For Coat, both \textit{Match} and \textit{Mismatch} groups achieve comparable scores, suggesting that the Judge consistently produces coherent and well-grounded explanations for its persuasive decisions.
For Kuairec with larger item space and noisier textual information, \textit{Match} rationales clearly outperform \textit{Mismatch} ones across all three criteria.
This indicates that, under a more complex task, the Judge concentrates its reasoning efforts on identifying logical connections that support \textit{Match} judgments.
In general, overall performance in both datasets demonstrates solid explainability in the judgment rationales, effectively addressing the absence of explainability in traditional offline evaluation. 




\begin{table}[t!]
    \centering
    \caption{Explainability analysis of the judgment rationales. Fleiss’ $\boldsymbol{\kappa}$ of each entry exceeds 0.75, indicating high inter-rater consistency.}
    \label{tab:human}
    \resizebox{\linewidth}{!}{
        \begin{tabular}{c c c c c}
            \toprule
            Dataset & Judgment & Coherence & Faithfulness & Persuasiveness  \\
            \midrule
            \multirow{3}*{Kuairec} 
                & Match &  84.67\%&  88.33\%&  86.67\% \\
                & Mismatch &69.33\%&  75.33\%&  72.33\%    \\
                & Overall &  77.00\%&  81.83\%& 79.50\%   \\
            \midrule
            \multirow{3}*{Coat} 
                & Match &  86.67\%&  81.33\%&  80.67\%   \\
                & Mismatch &84.00\% &85.33\%&  83.67\%    \\
                & Overall &  85.33\%& 83.33\% & 82.17\%  \\
            \bottomrule
        \end{tabular}
    }
\end{table}


\begin{figure}[t!]
    \centering  
    \subfigure[Match]{\includegraphics[width=0.495\linewidth]{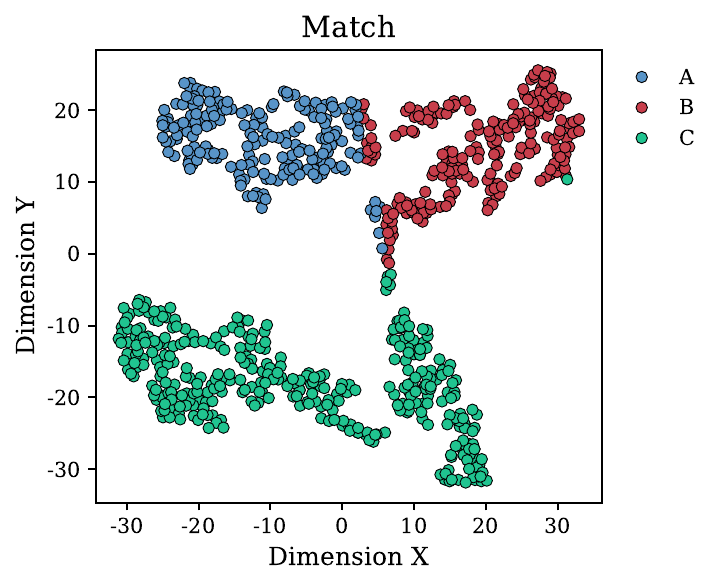}}
    \subfigure[Mismatch]{\includegraphics[width=0.495\linewidth]{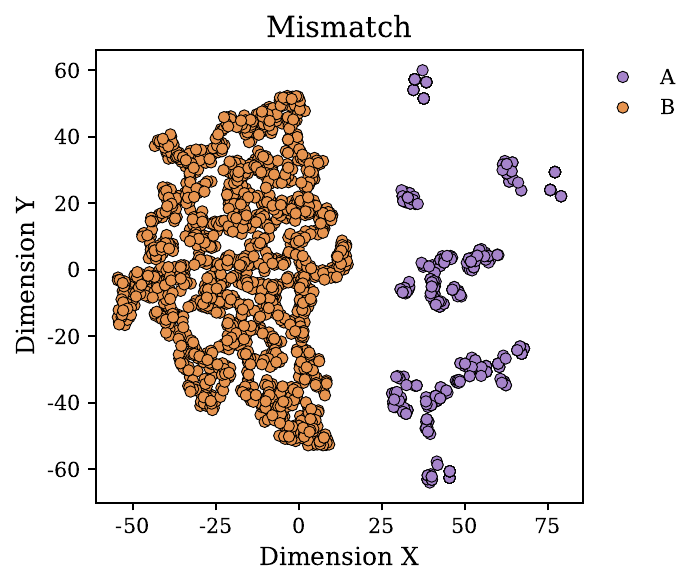}}
    \Description{match}
    \caption{Clustering results of the judgment rationales after Principal Component Analysis (PCA) dimension reduction.}
    \label{fig:match}
\end{figure}

\begin{table}[t]
    \centering
    \caption{Reason examples selected from the cluster centers.}
    \label{tab:reason}
    \resizebox{\linewidth}{!}{
    \begin{tabular}{c c p{0.72\linewidth}}
        \toprule
        Judgment & Cluster & Reason Example \\
        \midrule

        \multirow{3}{*}[-12.5ex]{Match}
            & A
            & \cellcolor{matchbg}The candidate jacket type is \textbf{packable}, which aligns with the user's established preference for \textbf{lightweight, functional outerwear}, indicating a strong stylistic and functional fit. \\[-2pt]
            \\[-6pt]

            & B
            & \cellcolor{matchbg}The candidate matches the user's established preference for \textbf{packable jackets} and \textbf{brown color}, supported by prior purchases of a packable jacket and multiple brown-colored items. \\[-2pt]
            \\[-6pt]

            & C
            & \cellcolor{matchbg}The candidate is a \textbf{black parka}, which aligns with the user's established preference for parkas (purchased in brown, gray, and other colors) and black (seen in insulated and motorcycle jackets). \\

        \midrule

        \multirow{2}{*}[-8ex]{Mismatch}
            & A
            & \cellcolor{mismatchbg}The candidate is for \textbf{men}, while the user's entire purchase history consists of \textbf{women's jackets}, indicating a gender mismatch. \\[-2pt]
            \\[-6pt]

            & B
            & \cellcolor{mismatchbg}The candidate is a track jacket in \textbf{gray}, which does not align with the user's strong preference for \textbf{black} and other specific colors like olive, green, and navy in similar jacket types, and gray is not represented in the history. \\
            
        \bottomrule
    \end{tabular}
    }
\end{table}

\begin{figure*}[t!]
    \centering 
    \subfigure[Recall@5]{
        \includegraphics[width=0.24\linewidth]{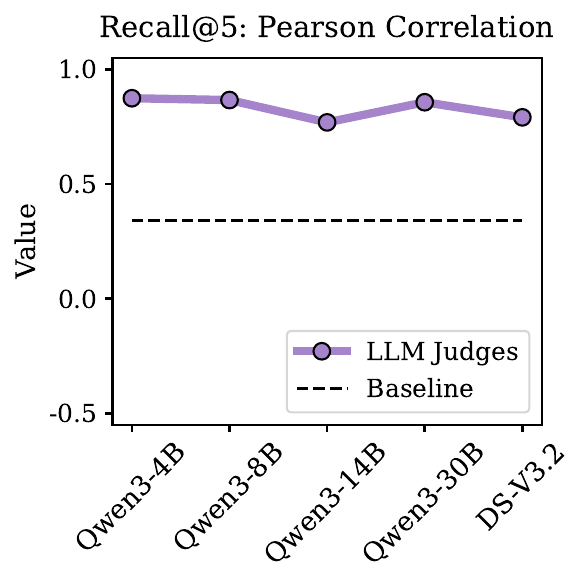}
        \includegraphics[width=0.24\linewidth]{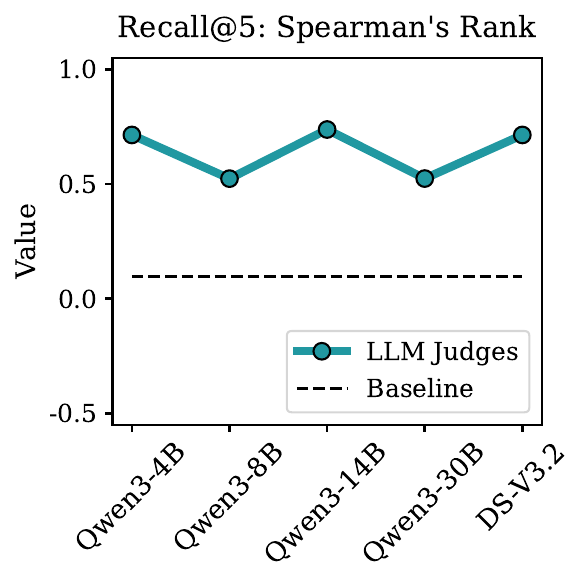}
        \includegraphics[width=0.24\linewidth]{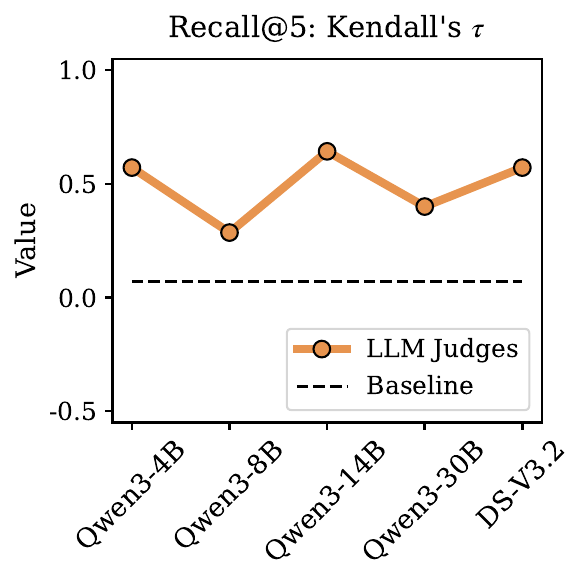}
        \includegraphics[width=0.24\linewidth]{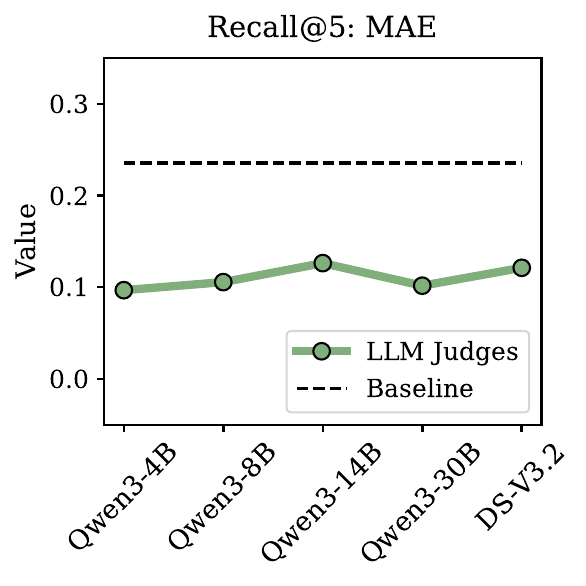}
    }

    \subfigure[NDCG@5]{
        \includegraphics[width=0.24\linewidth]{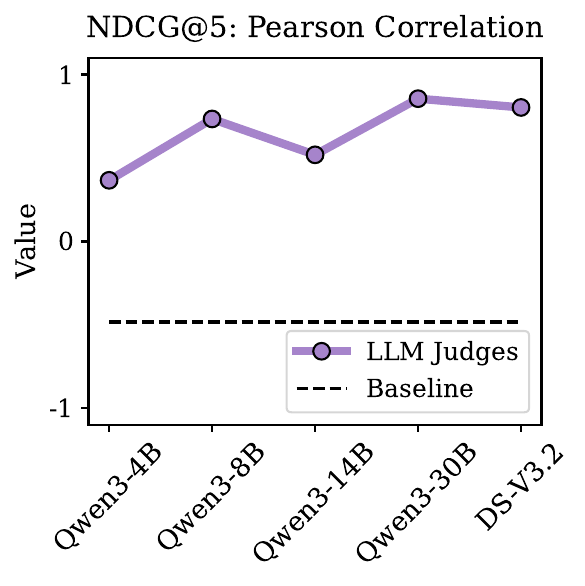}
        \includegraphics[width=0.24\linewidth]{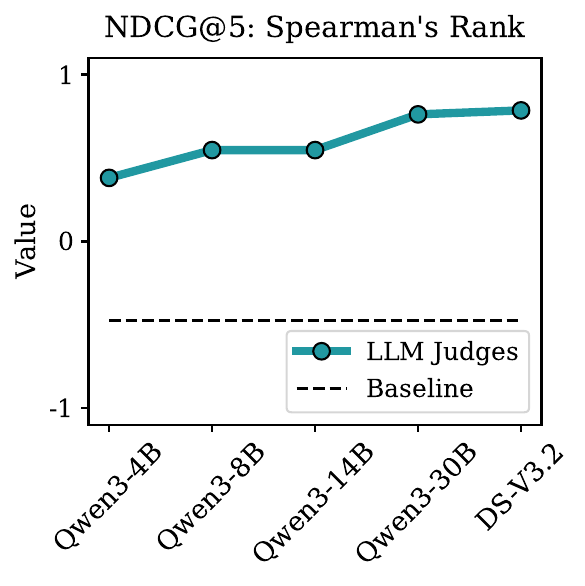}
        \includegraphics[width=0.24\linewidth]{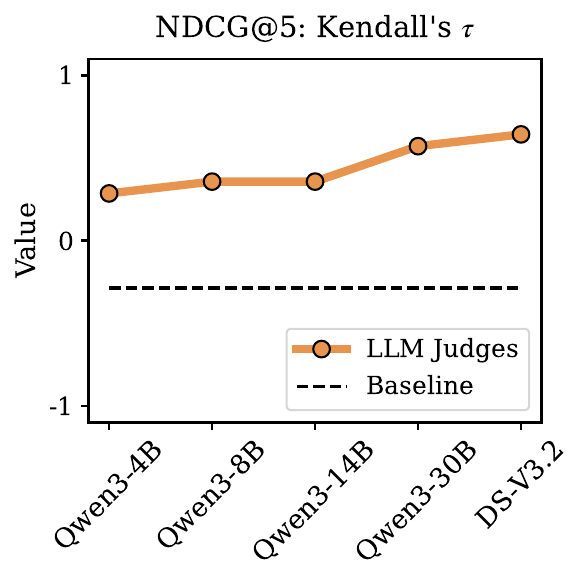}
        \includegraphics[width=0.24\linewidth]{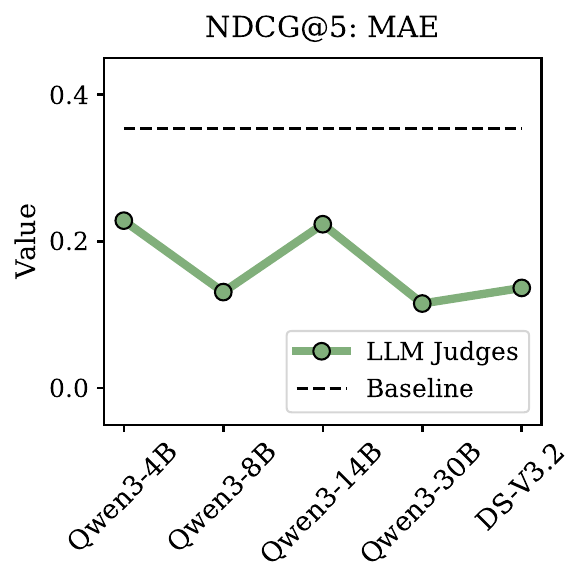}
    }
    \Description{robust}
    \caption{Evaluation performance on Recall@5 and NDCG@5 of the Coat dataset. Dashed baseline represents the benchmark of normal testing.}
    \label{fig:robust}
\end{figure*}

\emph{Case study of rationale patterns.}
We design the case study on Coat dataset to visualize the rationale landscape generated by the Judge. 
Specifically, we encode the rationales into sentence embeddings using the BGE-M3 tokenizer \cite{bge}.
Then we apply K-Means clustering \cite{kmeans} to the \textit{Match} and \textit{Mismatch} groups separately, and determine the optimal number of clusters using the elbow method.
The resulting clusters are visualized in Figure~\ref{fig:match}, where each cluster represents a distinct type of reasoning to the judgments.
We summarize the representative reason example that closest to each cluster center in Table~\ref{tab:reason}.
In the \textit{Match} group, there are three major categories of reasons.
Cluster A corresponds to logical extension, where the Judge generalizes user preferences to infer broader matches.
Cluster B reflects direct matching to support decisions through explicit attribute overlap.
Cluster C conforms to cross matching, where the Judge links elements in the candidate to multiple related signals in the history. 
For the \textit{Mismatch} group, Cluster A is primarily associated with gender discrepancies, while Cluster B covers various attribute-level conflicts with user preferences. 
Overall, these patterns further demonstrate the Judge's capability to produce structured and informative rationales, complementing the quantitative findings to establish comprehensive evaluation explainability.

\subsection{Study of Evaluation Robustness (RQ3)}
{\new To examine the evaluation robustness, we analyze the Judge sensitivity from two aspects.
First, we report the Judge performance under varying LLM backbone scales to validate inter-model robustness.
Second, we conduct ablation study on the input configuration, including both principle designs in prompt and volume of available input context.
We have the following observations.
}

{\new \emph{LLM scale analysis.}} Figure~\ref{fig:robust} summarizes the evaluation quality of each backbone in terms of Recall@5 and NDCG@5 on the Coat dataset. 
In each subfigure, the dashed horizontal line represents the normal testing benchmark, serving as a reference for consistency under the conventional offline evaluation setting.
Across both Recall and NDCG, the LLM Judge demonstrates strong consistency with the ground truth for all backbones and clearly surpasses the baseline benchmark.
In particular, correlation-based measures such as Pearson, Spearman, and Kendall remain significant across model scales, indicating that the Judge reliably preserves the relative ranking of models regardless of underlying LLM scales.
Additionally, the MAE remains consistently lower than the baseline, reflecting stable and superior score-level agreement with the unbiased results.

\begin{table}[t!] 
    \centering
    \caption{Ablation study of the Judge input configuration on the Coat dataset.}
    \label{tab:ablation}
    \setlength{\tabcolsep}{0.6mm}{
        \begin{tabular}{c c c c c c c}
            \toprule
            \multirow{2}*{Variant}& \multicolumn{3}{c}{Recall@5}& \multicolumn{3}{c}{NDCG@5}\\
            \cmidrule(lr){2-4}
            \cmidrule(lr){5-7}
            & Pearson& Spearman& Kendall& Pearson& Spearman&Kendall\\
            \midrule
            \textit{rewrite}& 88.36& 57.14& 42.86& 82.29& 71.82& 53.58\\
            \textit{w/o-RJ}& 90.74& 61.90& 50.00& 36.33& 38.10&28.57\\
            \textit{w/o-RS}& 70.37& 39.62& 30.88& 64.06& 64.29&42.86\\
            $\mathcal{L}$\textit{=0.6}& 82.85& 66.67& 50.00& 75.28& 57.14&28.57\\
            $\mathcal{L}$\textit{=0.3}& 76.91& 60.81& 54.29& 51.93& 54.76& 35.71\\
            $\mathcal{L}$\textit{=0.1}& 22.80& 43.11& 37.06& -34.62 &-33.33 &-21.43 \\
            \midrule
            Full& 85.70 & 52.41 & 40.01 & 85.61 & 76.19 & 57.14 \\
            \bottomrule
        \end{tabular}
    }
\end{table}

{\new \emph{Input configuration analysis.} As reported in Table~\ref{tab:ablation}, we experiment with different prompt designs through the first three variants, where the \textit{rewrite} variant paraphrases the prompt, while the \textit{w/o-RJ} and \textit{w/o-RS} variants remove corresponding reject-by-default and reasoning-then-scoring principles, respectively.
The \textit{rewrite} variant achieves performance comparable to the original Judge, showing that the framework is largely insensitive to surface-level wording.
The \textit{w/o-RJ} variant slightly improves Recall consistency by capturing interests in ambiguous items; however, this over-approval tendency introduces inevitable ranking noise and causes substantial degradation in NDCG.
The \textit{w/o-RS} variant underperforms the original Judge mainly due to the absence of the explicit reasoning step.
To further assess the stability of the Judge under varying input availability, we construct the remaining three  variants by randomly retaining 60\%, 30\%, and 10\% of input history length, denoted as $\mathcal{L}$\textit{=0.6}, \textit{0.3}, and \textit{0.1}. 
Results show that most metrics remain stable when at least 30\% of the history is retained, but completely fail at 10\%.
Since user histories in Coat are bounded within 14--16 items (\textit{cf.} Table~\ref{tab:appendix_input_stats}), the 10\% setting essentially represents a cold-start scenario, where even traditional holdout evaluation becomes unreliable due to insufficient test data.
Therefore, this confirms the robustness of the Judge under a reasonable range of history availability.
}


{\new In summary, the LLM Judge maintains evaluation robustness across backbone selection, prompt phrasing, and history availability. 
The robustness across varying backbone scales suggests that the framework does not exclusively depend on specific LLM capacity. 
The prompt ablation further demonstrates that the effectiveness of the Judge arises from our proposed structural principles rather than from surface-level wording. 
The analysis on input length reveals high tolerance of the Judge to various history sparsity levels.
Combined together, the LLM Judge framework provides a reliable proxy evaluation which can be flexibly deployed under different computational and data constraints.
}

%% file: related.tex
\section{Related Work}
\paragraph{Evaluation in recommender systems.} 
Existing approaches generally follow two paradigms: online and offline methods.
Online evaluation, such as A/B testing and interleaving experiments, directly measures user responses in real-world deployments, thus providing a faithful reflection of system utility \cite{evalcf, offon_eval}.
To mitigate the heavy dependence on live traffic, off-policy evaluation (OPE) is applied to estimate online performance using historical data sampled from the previously deployed logging policy \cite{ope, ope2}.
Nevertheless, in many research settings, real-time interaction environments and logging policies are often unavailable or costly to obtain.
Hence, offline evaluation remains the predominant paradigm for assessing recommender systems \cite{offline_eval_reliance, offline_rs_eval, offline_eval}.
In offline setting, substantial efforts have been devoted to addressing the problem of MNAR feedback.
For instance, studies in causal inference utilize propensity-weighting techniques to model item exposure, approximating a missing-at-random landscape for unbiased evaluation \cite{simpson, robust_eval, ips, ips2}.
However, this approach stays in the rigid ID matching manner to only focus on debiasing exposed items, while neglecting false negative samples.
Another direction lies in achieving online-offline alignment, which employs user dynamics from online experiments to guide offline metric design \cite{onoff_align, onoff_align2}.
Yet, as previously discussed, this method is largely constrained in applicability.

\paragraph{LLM-as-a-Judge.}
The emergence of LLM-as-a-Judge enables scalable, explainable, and human-aligned assessment across a wide range of tasks \cite{judgetask, judgetask2, judgetask3}.
This paradigm leverages the reasoning and world knowledge of LLMs to conduct nuanced judgments, often showing strong agreement with human evaluators \cite{laaj_survey}. 
In the field of information retrieval, LLM judges are initially investigated as a substitute for human annotators for relevance assessment \cite{llm_search, arena}.
Recently, this line of research has expanded into recommender systems.
For example, Zhang et al. \cite{llm_explan} employ LLM evaluators to assess explanations generated by explainable recommendation models, while Tokutake et al. \cite{Serendipity} introduce offline serendipity evaluation using LLMs.
These studies focus on specific aspects of recommendation, differing from our broader perspective.
Notably, Fabbri et al. \cite{podcast} propose a profile-based LLM judge for podcast recommendation evaluation.
However, their approach primarily relies on well-structured user podcast profiles, which restricts its all-scenario adaptability.
In contrast, our judge paradigm transforms user historical traces into semantic preference representations, ensuring strong generalizability across various recommendation tasks.

%% file: conclusion.tex
\section{Conclusion and Future Work}
In this paper, we present the LLM-as-a-Judge framework to offer reliable and explainable offline evaluation in Top-K recommendation.
To achieve reliable evaluation, we introduce a semantic proxy that summarizes user preferences from behavioral contexts, and replace rigid ID matching on holdout feedback with flexible semantic matching between the preference space and recommended items.
To realize explainable evaluation, we adopt a reasoning-then-scoring judging process to generate pointwise judgments together with informative rationales.
Our future work will investigate the extension of this evaluation paradigm beyond Top-K recommendation to other mainstream recommendation scenarios, such as sequential, cross-domain, and real-time recommendation.

%% file: appendix.tex
\section{Additional Details}

This appendix provides supplementary details that further clarify the proposed LLM Judge and support the reproducibility of the experimental results reported in the main paper.

\begin{table*}[htbp]
\centering
\caption{Field-level textual inputs used by the LLM Judge for each dataset. Only human-readable fields are used, and all fields are provided in full without truncation.}
\label{tab:appendix_field_protocol}
\begin{tabular}{l l l p{10cm}}
\toprule
Dataset &
Field / Attribute &
Type &
Description / Example \\
\midrule
\multirow{6}{*}{Kuairec}
& manual\_cover\_text
& Caption
& Cover text added by the video author (e.g., "Discovered by the little cutie") \\
& caption
& Caption
& Video title or description written by the author \\
& topic\_tag
& Caption
& Topic tags associated with the video (e.g., [Pomeranian Shun-suke, Live Streaming, Cute Pets Arrival]) \\
& first\_level\_category\_name
& Category
& First-level category name (e.g., \emph{Pet}) \\
& second\_level\_category\_name
& Category
& Second-level category name (e.g., \emph{Pet Daily}) \\
& third\_level\_category\_name
& Category
& Third-level category name (e.g., \emph{Pet Dog}) \\
\midrule
\multirow{3}{*}{Coat}
& gender
& Attribute
& Target gender of the jacket (e.g., women) \\
& jackettype
& Attribute
& Jacket type such as cropped, trench, fleece, or parka \\
& color
& Attribute
& Dominant color of the jacket (e.g., black, gray, green) \\
\bottomrule
\end{tabular}
\end{table*}

\begin{table*}[htbp]
\centering
\caption{Hyper-parameter settings for training the recommendation models.}
\label{tab:hyperparameter}

\begin{tabular}{c c  c c c c c c c  c  c}
\toprule
                  Dataset&Hyper-parameters& MF  & NeuMF    & NGCF     & LR-GCCF  & LightGCN &SimGCL&MGCCF&GTN&CAGED\\
\midrule 
  \multirow{5}*{Kuairec}&Batch Size& 4096& 4096& 4096& 4096& 4096& 4096& 4096& 4096 & 4096\\
  &Layer Number & 2&2 &2 & 2& 2& 2& 2&2 & 2 \\
  &Dimension Size &256 &256 &256 &256 &256 &256 &256 & 256 & 256 \\
  &Learning Rate & $1\times10^{-4}$& $5\times10^{-7}$&$5\times10^{-5}$ &$1\times10^{-5}$ &$1\times10^{-4}$ &$1\times10^{-4}$ &$1\times10^{-6}$ & $1\times10^{-4}$ &  $5\times10^{-4}$    \\
  &Regularization Rate & $1\times10^{-3}$& $0$&$5\times10^{-3}$ & $5\times10^{-3}$& $1\times10^{-3}$&$5\times10^{-3}$ &$5\times10^{-3}$ & $7\times10^{-3}$  &  $5\times10^{-3}$   \\
  
\midrule
\multirow{5}*{Coat}&Batch Size &512& 512& 512&512 & 512& 512& 512& 512& 512 \\
  &Layer Number & 2&2 &2 & 2& 2& 2& 2&2 & 2 \\
  &Dimension Size &256 &256 &256 &256 &256 &256 &256 & 256 & 256 \\
  &Learning Rate & $5\times10^{-4}$&$1\times10^{-5}$ & $5\times10^{-5}$&$1\times10^{-5}$ &$1\times10^{-4}$ &$1\times10^{-4}$ &$1\times10^{-5}$ & $1\times10^{-4}$ & $5\times10^{-4}$     \\
  &Regularization Rate  &$1\times10^{-2}$ &$0$ &$5\times10^{-3}$ & $5\times10^{-3}$&$5\times10^{-3}$ &$5\times10^{-3}$ &$5\times10^{-3}$ & $7\times10^{-3}$ & $5\times10^{-3}$     \\
\bottomrule
\end{tabular}
\end{table*}






\begin{table}[htbp]
\centering
\caption{Input statistics of user histories.}
\label{tab:appendix_input_stats}
\begin{tabular}{lcc}
\toprule
Statistic & Kuairec & Coat \\
\midrule
Number of Users& 1,411 & 290 \\
Average History Length& 49.24 & 15.51 \\
Median History Length& 50 & 16 \\
History Length Range& [26, 50] & [14, 16] \\
\midrule
Average Caption Length (Tokens)& 71.94 & 62.99 \\
Caption Length Range (Tokens)& [21, 2023] & [58, 69] \\
\bottomrule
\end{tabular}
\end{table}

\begin{table}[htbp]
\centering
\caption{Average runtime overhead of Qwen3-30B-A3B.}
\label{tab:appendix_overhead}
\begin{tabular}{cccc}
\toprule
Statistic/Call&Input Tokens&Output Tokens& Runtime\\
\midrule
Kuairec& 4130.6&51.7& 91.21ms\\
Coat& 494.6&43.0& 73.45ms\\
\bottomrule
\end{tabular}
\end{table}

\subsection{Input Context Details} \label{app_input}
To make the evaluation procedure fully transparent, we detail the exact textual inputs provided to the LLM Judge for each dataset, including field composition and history statistics.

Table~\ref{tab:appendix_field_protocol} summarizes the dataset-specific textual fields consumed by the LLM Judge. 
Across both datasets, the Judge operates exclusively on these human-readable textual information, which are provided in full with no truncation or heuristic compression applied.
For Kuairec, user histories and candidate items are represented using raw video-related textual metadata, including user-written captions, topic tags, and hierarchical category names. 
We preserve all such fields to reflect their inherent variability and semantic richness, representing a context-rich scenario.
For Coat, both user histories and candidate items are described using short and structured attributes, consisting of gender, jacket type, and color. 
This setting serves as a context-limited counterpart to Kuairec, enabling a controlled comparison at different levels of textual richness.


To further characterize the input context supplied to the Judge, we report statistics of user histories across both datasets in Table~\ref{tab:appendix_input_stats}.
The two datasets form complementary landscapes of history composition.
Kuairec represents the setting where user histories vary in length and are diverse in content, with substantial variation in both the number of interactions and the length of individual captions. 
In contrast, Coat has relatively consistent lengths in both user histories and captions. 
Therefore, we conduct the ablation study of history length based on this fixed-length structure, avoiding interference from length variability.


\subsection{Recommendation Model Details} \label{recmodel}
For reproducibility, we provide the hyper-parameter configurations used to train the recommendation models.
The trained models are then utilized to generate recommendation lists for evaluation under both traditional offline metrics and the LLM Judge.
The general hyper-parameter settings are summarized in Table~\ref{tab:hyperparameter},
while for other model-specific hyper-parameters, we follow the optimal settings reported in their original papers.

\subsection{Runtime Overhead Details} \label{overhead}
We report the average per-call overhead of Qwen3-30B-A3B deployed on 4 NVIDIA RTX A6000 GPUs in Table~\ref{tab:appendix_overhead}.
The input token count is mainly determined by the length and density of user histories, where Kuairec involves richer textual interactions and thus a substantially larger input volume than Coat.
In contrast, both the output token count and the per-call runtime remain consistently low on the two datasets.
This is because our pointwise reasoning-then-scoring scheme restricts each LLM call to a single user--candidate judgment, which keeps response generation lightweight and bounds the overall inference cost.

%% file: ref.bib
@article{offline_eval,
author = {Ca\~{n}amares, Roc\'{\i}o and Castells, Pablo and Moffat, Alistair},
title = {Offline evaluation options for recommender systems},
year = {2020},
issue_date = {Aug 2020},
publisher = {Kluwer Academic Publishers},
address = {USA},
volume = {23},
number = {4},
issn = {1386-4564},
journal = {Inf. Retr.},
month = mar,
pages = {387–410},
numpages = {24},
}

@inproceedings{podcast,
author = {Fabbri, Francesco and Penha, Gustavo and D'Amico, Edoardo and Wang, Alice and De Nadai, Marco and Doremus, Jackie and Gigioli, Paul and Damianou, Andreas and St\r{a}l, Oskar and Lalmas, Mounia},
title = {Evaluating Podcast Recommendations with Profile-Aware LLM-as-a-Judge},
year = {2025},
isbn = {9798400713644},
publisher = {Association for Computing Machinery},
address = {New York, NY, USA},
booktitle = {Proceedings of the Nineteenth ACM Conference on Recommender Systems},
pages = {1181–1186},
numpages = {6},
location = {
},
series = {RecSys '25}
}

@article{evalcf,
author = {Herlocker, Jonathan L. and Konstan, Joseph A. and Terveen, Loren G. and Riedl, John T.},
title = {Evaluating collaborative filtering recommender systems},
year = {2004},
issue_date = {January 2004},
publisher = {Association for Computing Machinery},
address = {New York, NY, USA},
volume = {22},
number = {1},
issn = {1046-8188},
journal = {ACM Trans. Inf. Syst.},
month = jan,
pages = {5–53},
numpages = {49},
}

@inproceedings{offon_eval,
author = {Beel, Joeran and Genzmehr, Marcel and Langer, Stefan and N\"{u}rnberger, Andreas and Gipp, Bela},
title = {A comparative analysis of offline and online evaluations and discussion of research paper recommender system evaluation},
year = {2013},
isbn = {9781450324656},
publisher = {Association for Computing Machinery},
address = {New York, NY, USA},
booktitle = {Proceedings of the International Workshop on Reproducibility and Replication in Recommender Systems Evaluation},
pages = {7–14},
numpages = {8},
location = {Hong Kong, China},
series = {RepSys '13}
}

@article{offline_eval_reliance,
author = {Bauer, Christine and Zangerle, Eva and Said, Alan},
title = {Exploring the Landscape of Recommender Systems Evaluation: Practices and Perspectives},
year = {2024},
issue_date = {March 2024},
publisher = {Association for Computing Machinery},
address = {New York, NY, USA},
volume = {2},
number = {1},
journal = {ACM Trans. Recomm. Syst.},
month = mar,
articleno = {11},
numpages = {31}
}

@inproceedings{closed_loop,
author = {Jadidinejad, Amir H. and Macdonald, Craig and Ounis, Iadh},
title = {Using Exploration to Alleviate Closed Loop Effects in Recommender Systems},
year = {2020},
isbn = {9781450380164},
publisher = {Association for Computing Machinery},
address = {New York, NY, USA},
booktitle = {Proceedings of the 43rd International ACM SIGIR Conference on Research and Development in Information Retrieval},
pages = {2025–2028},
numpages = {4},
location = {Virtual Event, China},
series = {SIGIR '20}
}

@article{simpson,
author = {Jadidinejad, Amir H. and Macdonald, Craig and Ounis, Iadh},
title = {The Simpson’s Paradox in the Offline Evaluation of Recommendation Systems},
year = {2021},
issue_date = {January 2022},
publisher = {Association for Computing Machinery},
address = {New York, NY, USA},
volume = {40},
number = {1},
issn = {1046-8188},
journal = {ACM Trans. Inf. Syst.},
month = sep,
articleno = {4},
numpages = {22}
}

@misc{robust_eval,
      title={Towards Robust Offline Evaluation: A Causal and Information Theoretic Framework for Debiasing Ranking Systems}, 
      author={Seyedeh Baharan Khatami and Sayan Chakraborty and Ruomeng Xu and Babak Salimi},
      year={2025},
      eprint={2504.03997},
      archivePrefix={arXiv},
}

@inproceedings{llm_rele_judge,
author = {Faggioli, Guglielmo and Dietz, Laura and Clarke, Charles L. A. and Demartini, Gianluca and Hagen, Matthias and Hauff, Claudia and Kando, Noriko and Kanoulas, Evangelos and Potthast, Martin and Stein, Benno and Wachsmuth, Henning},
title = {Perspectives on Large Language Models for Relevance Judgment},
year = {2023},
isbn = {9798400700736},
publisher = {Association for Computing Machinery},
address = {New York, NY, USA},
booktitle = {Proceedings of the 2023 ACM SIGIR International Conference on Theory of Information Retrieval},
pages = {39–50},
numpages = {12},
location = {Taipei, Taiwan},
series = {ICTIR '23}
}

@inproceedings{llm_explan,
author = {Zhang, Xiaoyu and Li, Yishan and Wang, Jiayin and Sun, Bowen and Ma, Weizhi and Sun, Peijie and Zhang, Min},
title = {Large Language Models as Evaluators for Recommendation Explanations},
year = {2024},
isbn = {9798400705052},
publisher = {Association for Computing Machinery},
address = {New York, NY, USA},
booktitle = {Proceedings of the 18th ACM Conference on Recommender Systems},
pages = {33–42},
numpages = {10},
location = {Bari, Italy},
series = {RecSys '24}
}

@misc{explan,
      title={Measuring "Why" in Recommender Systems: a Comprehensive Survey on the Evaluation of Explainable Recommendation}, 
      author={Xu Chen and Yongfeng Zhang and Ji-Rong Wen},
      year={2022},
      eprint={2202.06466},
      archivePrefix={arXiv},
}

@inproceedings{why,
author = {Figueiredo, Leticia Freire de and de A. Rocha, Antonio A. and Paes, Aline},
title = {Tell me why: how Explanation can affect Recommender Systems},
year = {2025},
isbn = {9798400713910},
publisher = {Association for Computing Machinery},
address = {New York, NY, USA},
booktitle = {Proceedings of the 2025 ACM International Conference on Interactive Media Experiences},
pages = {492–493},
numpages = {2},
location = {
},
series = {IMX '25}
}

@misc{laaj_survey,
      title={A Survey on LLM-as-a-Judge}, 
      author={Jiawei Gu and Xuhui Jiang and Zhichao Shi and Hexiang Tan and Xuehao Zhai and Chengjin Xu and Wei Li and Yinghan Shen and Shengjie Ma and Honghao Liu and Saizhuo Wang and Kun Zhang and Yuanzhuo Wang and Wen Gao and Lionel Ni and Jian Guo},
      year={2025},
      eprint={2411.15594},
      archivePrefix={arXiv},
}

@inproceedings{llm_search,
author = {Thomas, Paul and Spielman, Seth and Craswell, Nick and Mitra, Bhaskar},
title = {Large Language Models can Accurately Predict Searcher Preferences},
year = {2024},
isbn = {9798400704314},
publisher = {Association for Computing Machinery},
address = {New York, NY, USA},
booktitle = {Proceedings of the 47th International ACM SIGIR Conference on Research and Development in Information Retrieval},
pages = {1930–1940},
numpages = {11},
location = {Washington DC, USA},
series = {SIGIR '24}
}

@Inbook{holdout,
editor="Sammut, Claude
and Webb, Geoffrey I.",
title="Holdout Evaluation",
bookTitle="Encyclopedia of Machine Learning",
year="2010",
publisher="Springer US",
address="Boston, MA",
pages="506--507",
isbn="978-0-387-30164-8",
}

@article{eval_rs,
author = {Zangerle, Eva and Bauer, Christine},
title = {Evaluating Recommender Systems: Survey and Framework},
year = {2022},
issue_date = {August 2023},
publisher = {Association for Computing Machinery},
address = {New York, NY, USA},
volume = {55},
number = {8},
issn = {0360-0300},
journal = {ACM Comput. Surv.},
month = dec,
articleno = {170},
numpages = {38}
}

@article{offline_rs_eval,
author = {Castells, Pablo and Moffat, Alistair},
title = {Offline recommender system evaluation: Challenges and new directions},
journal = {AI Magazine},
volume = {43},
number = {2},
pages = {225-238},
year = {2022}
}

@inproceedings{ope,
author = {Jeunen, Olivier and Ustimenko, Aleksei},
title = {$\Delta$-OPE: Off-Policy Estimation with Pairs of Policies},
year = {2024},
isbn = {9798400705052},
publisher = {Association for Computing Machinery},
address = {New York, NY, USA},
booktitle = {Proceedings of the 18th ACM Conference on Recommender Systems},
pages = {878–883},
numpages = {6},
location = {Bari, Italy},
series = {RecSys '24}
}

@inproceedings{ope2,
author = {Narita, Yusuke and Yasui, Shota and Yata, Kohei},
title = {Debiased Off-Policy Evaluation for Recommendation Systems},
year = {2021},
isbn = {9781450384582},
publisher = {Association for Computing Machinery},
address = {New York, NY, USA},
booktitle = {Proceedings of the 15th ACM Conference on Recommender Systems},
pages = {372–379},
numpages = {8},
location = {Amsterdam, Netherlands},
series = {RecSys '21}
}

@inproceedings{onoff_align,
author = {Wilm, Timo and Normann, Philipp},
title = {Identifying Offline Metrics that Predict Online Impact: A Pragmatic Strategy for Real-World Recommender Systems},
year = {2025},
isbn = {9798400713644},
publisher = {Association for Computing Machinery},
address = {New York, NY, USA},
booktitle = {Proceedings of the Nineteenth ACM Conference on Recommender Systems},
pages = {967–970},
numpages = {4},
location = {
},
series = {RecSys '25}
}

@misc{onoff_align2,
      title={Bridging Offline-Online Evaluation with a Time-dependent and Popularity Bias-free Offline Metric for Recommenders}, 
      author={Petr Kasalický and Rodrigo Alves and Pavel Kordík},
      year={2023},
      eprint={2308.06885},
      archivePrefix={arXiv},
}

@inproceedings{ips,
author = {Yang, Longqi and Cui, Yin and Xuan, Yuan and Wang, Chenyang and Belongie, Serge and Estrin, Deborah},
title = {Unbiased offline recommender evaluation for missing-not-at-random implicit feedback},
year = {2018},
isbn = {9781450359016},
publisher = {Association for Computing Machinery},
address = {New York, NY, USA},
booktitle = {Proceedings of the 12th ACM Conference on Recommender Systems},
pages = {279–287},
numpages = {9},
location = {Vancouver, British Columbia, Canada},
series = {RecSys '18}
}

@inproceedings{Serendipity,
author = {Tokutake, Yu and Okamoto, Kazushi and Harada, Kei and Shibata, Atsushi and Karube, Koki},
title = {A Universal Framework for Offline Serendipity Evaluation in Recommender Systems via Large Language Models},
year = {2025},
isbn = {9798400720406},
publisher = {Association for Computing Machinery},
address = {New York, NY, USA},
booktitle = {Proceedings of the 34th ACM International Conference on Information and Knowledge Management},
pages = {5294–5298},
numpages = {5},
location = {Seoul, Republic of Korea},
series = {CIKM '25}
}

@article{judgetask,
title = {LLMs in medicine: The need for advanced evaluation systems for disruptive technologies},
journal = {The Innovation},
volume = {5},
number = {3},
pages = {},
year = {2024},
issn = {2666-6758},
author = {Yi-Da Tang and Er-Dan Dong and Wen Gao}
}

@article{judgetask2,
author = {Wang, Yilei and Zhao, Jiabao and Ones, Deniz and He, Liang and Xu, Xin},
year = {2025},
month = {01},
pages = {},
title = {Evaluating the ability of large language models to emulate personality},
volume = {15},
journal = {Scientific Reports},
}

@inproceedings{judgetask3,
 author = {Zheng, Lianmin and Chiang, Wei-Lin and Sheng, Ying and Zhuang, Siyuan and Wu, Zhanghao and Zhuang, Yonghao and Lin, Zi and Li, Zhuohan and Li, Dacheng and Xing, Eric and Zhang, Hao and Gonzalez, Joseph E and Stoica, Ion},
 booktitle = {Advances in Neural Information Processing Systems},
 pages = {46595--46623},
 publisher = {Curran Associates, Inc.},
 title = {Judging LLM-as-a-Judge with MT-Bench and Chatbot Arena},
 volume = {36},
 year = {2023}
}

@inproceedings{cf,
author = {Hu, Yifan and Koren, Yehuda and Volinsky, Chris},
title = {Collaborative Filtering for Implicit Feedback Datasets},
year = {2008},
isbn = {9780769535029},
publisher = {IEEE Computer Society},
address = {USA},
booktitle = {Proceedings of the 2008 Eighth IEEE International Conference on Data Mining},
pages = {263–272},
numpages = {10},
series = {ICDM '08}
}

@article{Hallucination,
author = {Huang, Lei and Yu, Weijiang and Ma, Weitao and Zhong, Weihong and Feng, Zhangyin and Wang, Haotian and Chen, Qianglong and Peng, Weihua and Feng, Xiaocheng and Qin, Bing and Liu, Ting},
title = {A Survey on Hallucination in Large Language Models: Principles, Taxonomy, Challenges, and Open Questions},
year = {2025},
issue_date = {March 2025},
publisher = {Association for Computing Machinery},
address = {New York, NY, USA},
volume = {43},
number = {2},
issn = {1046-8188},
journal = {ACM Trans. Inf. Syst.},
month = jan,
articleno = {42},
numpages = {55}
}

@inproceedings{ips2,
author = {Schnabel, Tobias and Swaminathan, Adith and Singh, Ashudeep and Chandak, Navin and Joachims, Thorsten},
title = {Recommendations as treatments: debiasing learning and evaluation},
year = {2016},
publisher = {JMLR.org},
booktitle = {Proceedings of the 33rd International Conference on International Conference on Machine Learning - Volume 48},
pages = {1670–1679},
numpages = {10},
location = {New York, NY, USA},
series = {ICML'16}
}

@inproceedings{kuairec,
author = {Gao, Chongming and Li, Shijun and Lei, Wenqiang and Chen, Jiawei and Li, Biao and Jiang, Peng and He, Xiangnan and Mao, Jiaxin and Chua, Tat-Seng},
title = {KuaiRec: A Fully-observed Dataset and Insights for Evaluating Recommender Systems},
year = {2022},
isbn = {9781450392365},
publisher = {Association for Computing Machinery},
address = {New York, NY, USA},
booktitle = {Proceedings of the 31st ACM International Conference on Information \& Knowledge Management},
pages = {540–550},
numpages = {11},
location = {Atlanta, GA, USA},
series = {CIKM '22}
}

@inproceedings{lightgcn,
author = {He, Xiangnan and Deng, Kuan and Wang, Xiang and Li, Yan and Zhang, YongDong and Wang, Meng},
title = {LightGCN: Simplifying and Powering Graph Convolution Network for Recommendation},
year = {2020},
isbn = {9781450380164},
publisher = {Association for Computing Machinery},
address = {New York, NY, USA},
booktitle = {Proceedings of the 43rd International ACM SIGIR Conference on Research and Development in Information Retrieval},
pages = {639–648},
numpages = {10},
location = {Virtual Event, China},
series = {SIGIR '20}
}

@ARTICLE{mf,
  author={Koren, Yehuda and Bell, Robert and Volinsky, Chris},
  journal={Computer}, 
  title={Matrix Factorization Techniques for Recommender Systems}, 
  year={2009},
  volume={42},
  number={8},
  pages={30-37},
}

@misc{lrgccf,
      title={Revisiting Graph based Collaborative Filtering: A Linear Residual Graph Convolutional Network Approach}, 
      author={Lei Chen and Le Wu and Richang Hong and Kun Zhang and Meng Wang},
      year={2020},
      eprint={2001.10167},
      archivePrefix={arXiv},
}

@inproceedings{ngcf,
author = {Wang, Xiang and He, Xiangnan and Wang, Meng and Feng, Fuli and Chua, Tat-Seng},
title = {Neural Graph Collaborative Filtering},
year = {2019},
isbn = {9781450361729},
publisher = {Association for Computing Machinery},
address = {New York, NY, USA},
booktitle = {Proceedings of the 42nd International ACM SIGIR Conference on Research and Development in Information Retrieval},
pages = {165–174},
numpages = {10},
location = {Paris, France},
series = {SIGIR'19}
}

@INPROCEEDINGS{mgccf,
  author={Sun, Jianing and Zhang, Yingxue and Ma, Chen and Coates, Mark and Guo, Huifeng and Tang, Ruiming and He, Xiuqiang},
  booktitle={2019 IEEE International Conference on Data Mining (ICDM)}, 
  title={Multi-graph Convolution Collaborative Filtering}, 
  year={2019},
  volume={},
  number={},
  pages={1306-1311},
}

@inproceedings{neumf,
author = {He, Xiangnan and Liao, Lizi and Zhang, Hanwang and Nie, Liqiang and Hu, Xia and Chua, Tat-Seng},
title = {Neural Collaborative Filtering},
year = {2017},
isbn = {9781450349130},
publisher = {International World Wide Web Conferences Steering Committee},
address = {Republic and Canton of Geneva, CHE},
booktitle = {Proceedings of the 26th International Conference on World Wide Web},
pages = {173–182},
numpages = {10},
location = {Perth, Australia},
series = {WWW '17}
}

@inproceedings{simgcl,
author = {Yu, Junliang and Yin, Hongzhi and Xia, Xin and Chen, Tong and Cui, Lizhen and Nguyen, Quoc Viet Hung},
title = {Are Graph Augmentations Necessary? Simple Graph Contrastive Learning for Recommendation},
year = {2022},
isbn = {9781450387323},
publisher = {Association for Computing Machinery},
address = {New York, NY, USA},
booktitle = {Proceedings of the 45th International ACM SIGIR Conference on Research and Development in Information Retrieval},
pages = {1294–1303},
numpages = {10},
location = {Madrid, Spain},
series = {SIGIR '22}
}

@inproceedings{gtn,
author = {Fan, Wenqi and Liu, Xiaorui and Jin, Wei and Zhao, Xiangyu and Tang, Jiliang and Li, Qing},
title = {Graph Trend Filtering Networks for Recommendation},
year = {2022},
isbn = {9781450387323},
publisher = {Association for Computing Machinery},
address = {New York, NY, USA},
booktitle = {Proceedings of the 45th International ACM SIGIR Conference on Research and Development in Information Retrieval},
pages = {112–121},
numpages = {10},
location = {Madrid, Spain},
series = {SIGIR '22}
}

@inproceedings{caged,
author = {Que, Yue and Zhang, Yingyi and Zhao, Xiangyu and Ma, Chen},
title = {Causality-aware Graph Aggregation Weight Estimator for Popularity Debiasing in Top-K Recommendation},
year = {2025},
isbn = {9798400720406},
publisher = {Association for Computing Machinery},
address = {New York, NY, USA},
booktitle = {Proceedings of the 34th ACM International Conference on Information and Knowledge Management},
pages = {2471–2481},
numpages = {11},
location = {Seoul, Republic of Korea},
series = {CIKM '25}
}

@misc{qwen,
      title={Qwen3 Technical Report}, 
      author={An Yang and Anfeng Li and Baosong Yang and others},
      year={2025},
      eprint={2505.09388},
      archivePrefix={arXiv},
}

@misc{ds,
      title={DeepSeek-V3 Technical Report}, 
      author={DeepSeek-AI and Aixin Liu and Bei Feng and others},
      year={2025},
      eprint={2412.19437},
      archivePrefix={arXiv},
}

@misc{bge,
      title={M3-Embedding: Multi-Linguality, Multi-Functionality, Multi-Granularity Text Embeddings Through Self-Knowledge Distillation}, 
      author={Jianlv Chen and Shitao Xiao and Peitian Zhang and Kun Luo and Defu Lian and Zheng Liu},
      year={2025},
      eprint={2402.03216},
      archivePrefix={arXiv},
}

@article{kmeans,
title = {K-means clustering algorithms: A comprehensive review, variants analysis, and advances in the era of big data},
journal = {Information Sciences},
volume = {622},
pages = {178-210},
year = {2023},
issn = {0020-0255},
author = {Abiodun M. Ikotun and Absalom E. Ezugwu and Laith Abualigah and Belal Abuhaija and Jia Heming},
}

@inproceedings{yahoo,
author = {Marlin, Benjamin M. and Zemel, Richard S.},
title = {Collaborative prediction and ranking with non-random missing data},
year = {2009},
isbn = {9781605584355},
publisher = {Association for Computing Machinery},
address = {New York, NY, USA},
booktitle = {Proceedings of the Third ACM Conference on Recommender Systems},
pages = {5–12},
numpages = {8},
location = {New York, New York, USA},
series = {RecSys '09}
}

@misc{arena,
      title={RecSys Arena: Pair-wise Recommender System Evaluation with Large Language Models}, 
      author={Zhuo Wu and Qinglin Jia and Chuhan Wu and Zhaocheng Du and Shuai Wang and Zan Wang and Zhenhua Dong},
      year={2024},
      eprint={2412.11068},
      archivePrefix={arXiv},
}

@inproceedings{kun1,
author = {Zhang, Xiaokun and Xu, Bo and Ren, Zhaochun and Wang, Xiaochen and Lin, Hongfei and Ma, Fenglong},
title = {Disentangling ID and Modality Effects for Session-based Recommendation},
year = {2024},
isbn = {9798400704314},
publisher = {Association for Computing Machinery},
address = {New York, NY, USA},
booktitle = {Proceedings of the 47th International ACM SIGIR Conference on Research and Development in Information Retrieval},
pages = {1883–1892},
numpages = {10},
location = {Washington DC, USA},
series = {SIGIR '24}
}

@inproceedings{kun2,
author = {Zhang, Xiaokun and Xu, Bo and Wu, Youlin and Zhong, Yuan and Lin, Hongfei and Ma, Fenglong},
title = {FineRec: Exploring Fine-grained Sequential Recommendation},
year = {2024},
isbn = {9798400704314},
publisher = {Association for Computing Machinery},
address = {New York, NY, USA},
booktitle = {Proceedings of the 47th International ACM SIGIR Conference on Research and Development in Information Retrieval},
pages = {1599–1608},
numpages = {10},
location = {Washington DC, USA},
series = {SIGIR '24}
}

@ARTICLE{kun3,
  author={Zhang, Xiaokun and Xu, Bo and Li, Chenliang and He, Bowei and Lin, Hongfei and Ma, Chen and Ma, Fenglong},
  journal={IEEE Transactions on Knowledge and Data Engineering}, 
  title={A Survey on Side Information-Driven Session-Based Recommendation: From a Data-Centric Perspective}, 
  year={2025},
  volume={37},
  number={8},
  pages={4411-4431},
}
